\documentclass[nomathfonts]{aipproc} 
\layoutstyle{6s}
\usepackage{xspace,amsmath,amssymb,bbm}
\input{alphabet}   
\input{abrege}    
\def\pth#1{\left(#1\right)}                
\def\acc#1{\left\{#1\right\}}              
\def\cro#1{\left[#1\right]}

\def\esp{{\mathrm{E}}\,}					 
\def\var{{\mathrm{var}}\,}

\newsavebox{\fminibox}
\newlength{\fminilength}
\newenvironment{fminipage}[1][\linewidth]
  {\setlength{\fminilength}{#1}
   \begin{lrbox}{\fminibox}\begin{minipage}{\fminilength}}
  {\end{minipage}\end{lrbox}\noindent\fbox{\usebox{\fminibox}}}

  \def\+{^\dagger}

\def\nequiv{\not\kern-.05em\equiv}
\def\egal{\kern-.5em=\kern-.5em}        
\def\propt{\kern-.2em\propto\kern-.2em} 
\def\wh#1{\widehat{#1}}                 

\def\argmax{\mathop{\mathrm{arg\,max}}} 
\def\argmin{\mathop{\mathrm{arg\,min}}} 

\def\intdouble{\int\kern-0.3em\int}
\def\inttriple{\int\kern-0.3em\int\kern-0.3em\int}

\def\rond#1{\overset{\kern-0.33em~_\circ}{#1}}
\def\rondit[#1]#2{\overset{\kern#1~_\circ}{#2}}

\def\babs{\begin{abstract}}             \def\eabs{\end{abstract}}
\def\barr{\begin{array}}                \def\earr{\end{array}}
\def\bcc{\begin{center}}                \def\ecc{\end{center}}

\def\bdes{\begin{description}}          \def\edes{\end{description}}
\def\bdoc{\begin{document}}             \def\edoc{\end{document}}
\def\ben{\begin{enumerate}}             \def\een{\end{enumerate}}
\def\beqn{\begin{eqnarray}}             \def\eeqn{\end{eqnarray}}
\def\beqnl#1{\beqn\label{#1}}           \def\eeqnl#1{\label{#1}\eeqn}
\def\beqnx{\begin{eqnarray*}}           \def\eeqnx{\end{eqnarray*}}
\def\bseqn{\begin{subeqnarray}}         \def\eseqn{\end{subeqnarray}}
\def\beq#1\eeq{\begin{equation}#1\end{equation}}
\def\bal#1\eal{\begin{align}#1\end{align}}
\def\balx#1\ealx{\begin{align*}#1\end{align*}}
\def\beqx{$$}                           \def\eeqx{$$}
\def\bfig{\protect\begin{figure}}       \def\efig{\protect\end{figure}}
\def\bfigx{\protect\begin{figure*}}     \def\efigx{\protect\end{figure*}}
\def\bfigt{\protect\begin{figurette}}   \def\efigt{\protect\end{figurette}}
\def\bfl{\begin{flushleft}}             \def\efl{\end{flushleft}}
\def\bfr{\begin{flushright}}            \def\efr{\end{flushright}}
\def\bit{\begin{itemize}}               \def\eit{\end{itemize}}
\def\bmi{\begin{minipage}}              \def\emi{\end{minipage}}
\def\bfmi{\begin{fminipage}}            \def\efmi{\end{fminipage}}
\def\bpic{\begin{picture}}              \def\epic{\end{picture}}
\def\bqu{\begin{quote}}                 \def\equ{\end{quote}}
\def\bqun{\begin{quotation}}            \def\equn{\end{quotation}}
\def\bsl{\begin{slide}}                 \def\esl{\end{slide}}
\def\btabb{\begin{tabbing}}             \def\etabb{\end{tabbing}}
\def\btabl{\begin{table}}               \def\etabl{\end{table}}
\def\btablx{\begin{table*}}             \def\etablx{\end{table*}}
\def\btab{\begin{tabular}} 
\def\btabu{\begin{tabular}}             \def\etabu{\end{tabular}}
\def\btabx{\begin{tabular*}}            \def\etabx{\end{tabular*}}
\def\bbib{\begin{thebibliography}{999}} \def\ebib{\end{thebibliography}}
\def\bver{\begin{verbatim}}             \def\ever{\end{verbatim}}
\def\bca{\begin{cases}}                          \def\eca{\end{cases}}
   
\def\fh{\wh{f}}
\def\qh{\wh{q}}
\def\fbh{\wh{\fb}}
\def\hbh{\wh{\hb}}
\def\qbh{\wh{\qb}}
\def\thetabh{\wh{\thetab}}
\def\sbh{\wh{\sb}}
\def\mubh{\wh{\mub}}
\def\lambdabh{\wh{\lambdab}}
\def\lambdah{\wh{\lambda}}
\def\nubh{\wh{\nub}}
\def\nbh{\wh{\nb}}
\def\nuh{\wh{\nu}}
\def\thetah{\wh{\theta}}
\def\gbh{\wh{\gb}}
\def\onebx#1#2{{\bf 1}_{{#1}\times{#2}}}
\def\oneb{{\bf 1}}
\def\stack#1#2{\barr{@{}c@{}} {#1}\\ {#2}\earr}
\def\mode#1{\mbox{mode}\{#1\}}
\def\esp#1{\mbox{E}\{#1\}}
\def\var#1{\mbox{Var}\{#1\}}
\def\espx#1#2{\mbox{E}_{#1}\{#2\}}
\def\expf#1{\exp\left\{#1\right\}}
\def\pmatrix#1{\left[\begin{array}{cc} #1 \end{array}\right]}
\def\intg{\int\kern-1.1em\int}
\def\intd{\int\kern-.8em\int}
\def\lra{\longrightarrow}
\def\argmax#1#2{\mbox{arg}\max_{#1}\left\{#2\right\}}
\def\argmin#1#2{\mbox{arg}\min_{#1}\left\{#2\right\}}
\def\d#1{\,\mbox{d}#1}
\def\disp{\displaystyle}
\def\itema{\smallskip\item}

\title{Yet Another Analysis of Dice Problems.
\thanks{To appear in Proceedings of American Institute of Physics: 
Proceedings of MaxEnt2002, the 22nd International Workshop on Bayesian 
and Maximum Entropy methods (Aug. 3-9, 2002, Moscow, Idaho, USA).}
}

\author{Ali Mohammad-Djafari}
 {
  address = {Laboratoire des Signaux et Syst\`emes,\linebreak 
  Unit\'e mixte de recherche 8506 (CNRS-Sup\'elec-UPS) \linebreak  
  Sup\'elec, Plateau de Moulon, 91192 Gif-sur-Yvette, France},
  email = {djafari@lss.supelec.fr}
 }

\begin{document}
\begin{abstract}
During the MaxEnt 2002 workshop in Moscow, Idaho, Tony Vignaux 
asked again a few simple 
questions about using Maximum Entropy or Bayesian approaches for 
the famous Dice problems which have been analyzed many times through 
this workshop and also in other places. 
Here, there is another analysis of these problems. 
I hope that, this paper will answer a few questions of Tony and other 
participants of the workshop on the situations where we can use 
Maximum Entropy or Bayesian approaches or even the cases where we can 
actually use both of them. 

\bigskip\noindent{\bf keywords.~}  
Dice problems and probability theory, Maximum Likelihood, 
Bayesian inference, Maximum A Posteriori, Entropy, 
Maximum entropy, Maximum entropy in the mean. 
\end{abstract}

\maketitle

\section{Introduction}
\label{Introduction}

Dice problems have been analyzed many times 
(See mainly Ed. Jaynes papers \cite{Jaynes68,Jaynes78,Jaynes82,Jaynes85} 
and also \cite{Frieden85b,Frieden87,Shore80,VanCampenhout81}),  
but it seems that still many questions are open. 
In this note, I will try to answer some of them. 
Before starting, we need to set up precise notation and describe precisely 
the context. 

Let's consider an imaginary die with $K$ faces ($K=6$ 
is the ordinary die) where on each face there is a number. 
We note these numbers $\gb=[g_1,\ldots,g_{K'}]$.  
$K$ is the number of elementary states and commonly, $K'=K$ 
and $g_k=k$, but we may also consider the 
cases where $g_k$ are any other numbers (integer or real) distinct or not.  

Let's also represent by $X$ the variable corresponding to 
face number and by $G$ 
the variable corresponding to the number written on the faces. 
So, $X$ may take values $\acc{1,\ldots,K}$ and $G$ can take values 
$\acc{g_1,\ldots,g_{K'}}$. 
Then, we can define $P(X=k)$ and $P(G=g_k)$. 
If the $g_k$ are distinct numbers, \ie, $K=K'$, they are equal 
$P(X=k)=P(G=g_k)=\theta_k$, but 
note that $\esp{X}=\sum_k k \theta_k \not=\esp{G}=\sum_k g_k \theta_k$. 
If $g_k$ is a monotone function of $k$, then it is easy to relate 
$\esp{X}$ to $\esp{G}$, but it may not always be the case. 

Note also that, in many dice problems, the main hypothesis is that they are 
fair. Then assigning the probability distributions becomes a combinatorial 
computation. For example, suppose we throw two dice and count the sums $S$ 
of the two faces numbers. 
We want to assign the probabilities  $p_j=P(S=s_j)$. 

First, we assume $g_k=k$ and note that $S$ can take the 
values in the set $\Omega=\{2,3,\ldots,12\}$ and $|Q|=11$. 
We must be careful here because the event $S=s_j$ can occur 
$q(s_j)=6-|7-s_j|$ times. 
For example, 
$S=2$ occurs one time $E_j=\{(1,1)\}$, but $S=5$ occurs 5 times  $E_j=\{(1,4),(2,3),(3,2),(4,1)\}$. Now, using the basic principle of 
\emph{equal weight} of statistical mechanics or \emph{insufficient reason} 
of Laplace, we assign $p_j=P(S=s_j)\propto |E_j|$ which gives 
$p_j=P(S=s_j)=q(s_j)/\sum_{j=1}^{|Q|} q(s_j)$. 

In a more general case, we may have $L$ dice and may want to define 
the events such that  
$E_j=\{(X_1=x_{1},\ldots,X_L=x_{L})\}$ or 
$E_j=\{(X_1=x_{1},\ldots,X_L=x_{L}) ~:~ \sum_l x_l=s_j\}$ and assign 
them probabilities. 
We may also consider the case where we throw $L$ dice simultaneously 
$N$ times which is not the same as throwing $N$ dice simultaneously 
$L$ times, except the case where the dice are identical.   
We may also consider the cases where the number of throwing the dice 
are different, \ie, the dice $l$ has been thrown $N_l$ times. 

In some other analysis, we may not know if the die is loaded 
or not. This may be 
one of the questions to be answered. To be able to answer to a question, 
we may need to gather relevant data. These data may be of different 
form and thus, as we will see in the following, the way to use them to 
answer a question may also differ. 

Before gathering any data, we may define the \emph{question} to 
be answered. For example, if we want to know if the die is loaded or not, 
we may be interested to infer about $\thetab$. 
Also, before gathering any data, we may make 
hypotheses and we may be able to translate the knowledge contained in 
these hypotheses by an \aprio probability law $\pi(\thetab)$. 

For example, we 
may assume that the die is not loaded and assume  $\theta_1=\theta_2=\ldots=\theta_K=1/K$ or 
choose a uniform prior for $\pi(\thetab)$ over the set 
$\{\thetab~:~ \theta_k\in[0,1] \& \sum_k \theta_k=1\}$. 
Note that, even if they translate to a common-sounding hypothesis, mathematically speaking, they are not exactly the same. 
The former says 
$P(\sum_k\theta_k\not=1)=0$ and  
$P(\theta_k\not=\theta_l, k\not=l)=0$ and 
$P(a<\theta_k\le b)=(b-a), \; \forall 1>b>a>0$. 

We may also be able to 
associate a likelihood function $P(D|\thetab)$ with the data to represent 
the amount of knowledge about the unknown parameters contained in the data. 
We will see however that this may not be easy in some cases. 

The questions may also be different: We may want to know if the die is 
loaded or not 
or we may want to know what is the probability that the next face be 
the face $k$, or still, what are the numbers written on the faces of 
the die.   

Let us start by a simple and easy problem which, here after, 
we call Problem 1. 

\section{Problem 1}

We have observed the complete data $\xb=[x_1,\ldots,x_N]$ and we know 
the number of states $K$ (number of faces). The question is to estimate 
$\thetab=[\theta_1,\ldots,\theta_K]$ where $\theta_k=P(X=k)$ 
is the probability of the event face $k$ up. 

Here is a Matlab program which simulates this data generation:
\\~\\ 
{\tt K=6;N=100;x=round((K-1)*rand(N,1))+1;}  
\\~\\ 
and the following is an example (an $N$ sample) of this data set:
\\~\\
$\xb=[
4,2,2,2,1,5,4,5,1,4,3,3,6,6,4,6,6,4,4,1,1,2,1,6,4,2,4,2,3,2,2,6,2,2,1,6,\\
5,5,6,3,5,4,2,2,4,4,4,3,6,6,4,5,2,5,3,5,2,5,1,3,3,4,3,1,3,3,5,3,3,2,5,5,3,4,4,\\
3,3,3,4,1,2,4,4,5,4,5,6,5,6,5,5,5,1,1,4,1,5,2,1,6]$.
\\~\\
Note that, if we re-run the program, we obtain a different data set. 
Here are the results of a second run:
\\~\\ 
$\xb=[
6,2,4,3,5,5,3,1,5,3,4,5,6,5,2,3,6,6,3,5,1,3,5,1,2,2,2,4,2,2,1,5,3,6,3,3,\\ 
5,4,2,4,5,1,4,3,5,4,5,3,3,2,2,4,3,4,2,4,3,5,5,4,3,5,5,4,5,4,3,2,3,4,5,3,5,4,3,\\ 5,4,3,4,4,5,6,4,5,2,6,2,2,5,5,2,1,5,2,2,4,2,3,1,6]$.
\\~\\
These two data sets can represent two different experiences using 
the same die. 

Let $n_k$ denote the number of times the face $k$ has shown up 
$n_k=\#(X=k)$. Then we have $\sum_k n_k=N$. 
Here is a Matlab program which computes these numbers:\\~\\  
{\tt 
nk=zeros(K,1); \\ 
for k=1:K \\
\hspace*{5mm} nk(k)=sum(x==k); \\
end 
}
\\~\\  
and here are the results for the two above data sets:\\~\\
Data set 1:\quad $\nb=[13,17,17,21,19,13]$ and \\  
Data set 2:\quad $\nb=[07,19,21,20,25,08]$.\\  
  
\bigskip 
Now, let's start by asking about the values of $\theta_k$. If all the 
$\theta_k$ are the same value, we can say that the die is not loaded, 
but if they are too different from each other, we may say that the 
die is loaded. 

A wise man can say:  This is an easy problem. If each trial has been 
done identically and independently, then it is reasonable to 
\emph{estimate} each $\theta_k$ by $\theta_k=n_k/N$ and no need 
for more complex mathematics. 
But if we ask: How \emph{confident} or (how sure) are you about 
these values? He may say: hum..., let's use the probability theory. 

Assume we know $K$ and we have given $\xb$ 
(and thus we now $N$) and assume that the die has been thrown  
always in the same manner and independently.   
Here then, we can write the complete likelihood function
\beq
P(\xb|\thetab) = 
\prod_k C_N^{n_k}\, \theta_k^{n_k}\, (1-\theta_k)^{N-n_k}.
\eeq
Note that in the right hand side of this expression, $\xb$ is present 
through $n_k$ and we can write $P(\xb|\thetab)=P(\nb|\thetab)$. 

Then the likelihood $\Lc(\thetab)=P(\xb|\thetab)$ and we have  
\beq
\ln \Lc(\thetab)= \sum_k \cro{ {n_k}\ln \theta_k + (N-n_k)\ln(1-\theta_k) } +c(n_k,N)
\eeq
where $c(n_k,N)=\sum_k \ln C_N^{n_k}$ does not depend on $\thetab$. 

Knowing that each parameter $\theta_k\in[0,1]$, we can choose a 
uniform prior $\pi(\theta_k)=1$ on this interval. However, we know that 
$\sum_k \theta_k=1$, then we can define the set 
$\Theta=\{\thetab~:~ \theta_k\in[0,1] \& \sum_k \theta_k=1\}$ 
and thus define a uniform prior on this set 
$\pi(\thetab)=1, \; \forall \thetab\in\Theta$ 
and zero elsewhere, and thus obtain the 
\apost law 
\beq
\pi(\thetab|\xb)=\frac{\Lc(\thetab) \, \pi(\thetab)}{m(\xb)} 
=\frac{1}{m(\xb)} 
\prod_k C_N^{n_k}\, \theta_k^{n_k}\, (1-\theta_k)^{N-n_k}
\eeq
which is defined on the same set $\Theta$ and  
where $m(\xb)$ is the marginal or evidence function:
\beq
m(\xb)=\intg_{\Theta} \Lc(\thetab) \, \pi(\thetab) \d{\thetab}
=\intg_{\Theta} \d{\thetab} \prod_k C_N^{n_k}\, \theta_k^{n_k}\, (1-\theta_k)^{N-n_k}. 
\eeq
Thus, we have
\beq
\ln \pi(\thetab|\xb)=  
\sum_k \cro{ {n_k}\ln \theta_k + (N-n_k)\ln(1-\theta_k)} -\ln m(\xb). 
\eeq

Now, if we are only interested by the value of $\thetabh^{MAP}$ which 
has the highest probability, we can compute it by 
putting the derivative of ~$\ln \pi(\thetab|\xb)$ with respect to each 
parameter $\theta_k$ to obtain
\beq
\partial \ln \pi(\thetab|\xb) / \partial \theta_k=
\frac{n_k}{\theta_k} - \frac{N-n_k}{1-\theta_k} = \frac{n_k-N \theta_k}{\theta_k(1-\theta_k)}=0 
\lra \thetah^{MAP}_k=\frac{n_k}{N}.
\eeq
There is only one possible solution to this equation and there is not 
any ambiguity. 
Here are the results for the two above data sets:\\~\\  
Data set 1: \quad 
$\thetabh^{MAP}=[0.1300, 0.1700, 0.1700, 0.2100, 0.1900, 0.1300]$~~ 
and \\ 
Data set 2: \quad 
$\thetabh^{MAP}=[0.0700,0.1900,0.2100,0.2000,0.2500,0.0800]$. 

But, we must be careful here on the interpretations that we can give 
to these numerical values. We may want to answer the following questions: 
\bit
\item Do these two data sets come from the same die? 
\item Is this die loaded? 
\item What is the probability of seeing face $k$ up based on the data 
set 1 or the data set 2? 
\item If I throw this die $100$ times again, what will be the number of 
times I will see face $k$ up? 
\eit
We have still too much to do before being able to give correct answers  
these questions. 
                                                              
\section{Problem 2}

Assume now that, in place of $\xb$, we have only access to the data 
$\nb=(n_1,\ldots,n_K)$ and know the values of $K$ and $N$ (or if we knew 
that $\sum_k n_k=N$). 
It is easy to see that we obtain exactly 
the same result, because $(\nb, K, \sum_{k=1}^K n_k=N)$ define perfectly 
the likelihood and form sufficient statistics about this problem. 

\medskip 
Note however that, in both cases, the likelihood $\Lc(\thetab)$ 
is not defined for $\theta_k=0$ and $\theta_k=1$ and consequently, 
the posterior pdf $\pi(\thetab|\nb)$ may 
not be a proper pdf. We are going to analyze properly this point. 

First, noting that the likelihood function in the previous section 
$\Lc(\thetab)= \prod_k l(\theta_k)$ and 
$\pi(\thetab)=\prod_k \pi(\theta_k)$, 
we also have 
$\pi(\thetab|\xb)=\prod_k \pi(\theta_k|\xb)=\prod_k \pi(\theta_k|n_k)$.  
Thus, we can work hereafter only with the functions $l(\theta)$, 
$\pi(\theta)$, $\pi(\theta|x)$ and $m(x)$ which is given by 
\cite{Robert92a,Robert97b} 
\beq
m(x)=\int_0^1 \pi(\theta) \, \theta^x \,  (1-\theta)^{(N-x)} \d{\theta}.
\eeq
With a uniform prior $\pi(\theta)=1$ we have 
\beq
m(x)=\int_0^1  
\theta^x \,  (1-\theta)^{(N-x)} \d{\theta}=\Bc(x+1, N-x+1)
\eeq
where $\Bc(\alpha, \beta)$ is the Beta probability density function (pdf) 
\beq
f(x|\alpha,\beta)=\frac{1}{B(\alpha,\beta)} x^{\alpha-1} (1-x)^{(\beta-1)} 
\eeq
which is defined for $\alpha>0$, $\beta>0$ and $x\in[0,1]$ and where  
\beq
B(\alpha,\beta)=\int_0^1 x^{\alpha-1} (1-x)^{(\beta-1)} \d{x} 
\eeq
and we have:  
\beqnx
\mode{x}=\frac{\alpha-1}{\alpha+\beta-2}, \quad  
\esp{x}=\frac{\alpha}{\alpha+\beta} \quad \mbox{and} \quad 
\var{x}=\frac{\alpha\beta}{(\alpha+\beta)^2(\alpha+\beta+1)}.
\eeqnx 

Consequently, the posterior law, whose expression is
$\pi(\thetab|\nb)=\Bc(\nb, N-\nb)$ or equivalently 
$\pi(\theta_k|n_k)=\Bc(n_k, N-n_k)$, is only bounded if $N>n_k>0$. 
The MAP estimators $\thetah_k^{MAP}=\frac{n_k-1}{N-2}$ 
do not exist if $n_k<1$ or if $n_k>N-1$ and if $N\le 2$.  

The posterior mean estimators  
$\thetah_k^{PM}=\frac{n_k}{N}$ exist if $N>0$ and 
the posterior variances 
$\var{\theta}_k=\frac{(n_k)(N-n_k)}{(N^2(N+1)}$ 
exist if $0<n_k<N$ and $N>0$. 
Note also that when $n_k=1$ the corresponding MAP estimator 
is $\theta_k=0$ and  
when $n_k=N-1$ the corresponding MAP estimator is $\theta_k=1$. 
This shows a kind of  
bias of the estimator toward $\theta_k=0$ and $\theta_k=1$ (See Table 1). 

One may want to have a proper posterior law $\pi(\thetab|\nb)$ for the 
whole range of possible values of the parameters $\theta_k\in[0,1]$ 
and the data $n_k=[0,1,\ldots,N]$. 
This can be done via other choices for the prior law. 
In the two previous cases, we choose a uniform \aprio for $\theta_k$. 
Some authors 
argued that this choice is too biased against extreme 
values $0$ and $1$ and proposed to use 
\beq
\pi(\theta_k)=[\theta_k (1-\theta_k)]^{-1}=\theta_k^{-1} (1-\theta_k)^{-1}.
\eeq
\medskip 
Note also that, again with this prior, the normalization factor or 
the evidence function $m(x)$ is given by
\beq
m(x)=\int_0^1 [\theta (1-\theta)]^{-1}\, 
 \theta^x \,  (1-\theta)^{(N-x)} \d{\theta}=
\Bc(x+1, N-x+1)
\eeq
which yields $\pi(\thetab|\nb)=\Bc(\nb+1, N-\nb+1)$ which is bounded if 
$N-1>n_k>0$ (See Table 1). 

A more general choice is 
\beq
\pi(\theta_k)=\theta_k^{a-1} (1-\theta_k)^{b-1} 
\eeq
which results to 
\beq
m(x)=\int_0^1 \theta^{a-1} (1-\theta)^{b-1}\, 
\theta^x \,  (1-\theta)^{(N-x)} \d{\theta}
=\Bc(x+a, N+b-x)
\eeq
which result to $\pi(\thetab|\nb)=\Bc(\nb+a, N+b-\nb)$   
which is bounded if $N-b>n_k>1-a$. 
Then, the mean values $\theta_k=(n_k+a)/(N+b+a)$ have  
the limit value $\theta_k=n_k/N$ when $a=b\mapsto 0$. 
The following Table summarizes these points. 

\bigskip 
\begin{table}[htb]
\btabu{|c|c|c|c|c|c|}\hline\hline  
~~~~ $\alpha>0$~~~~ & 
~~~~ $\beta>0$~~~~   & 
~~~~ $\alpha+\beta$~~~~ & 
~~~~ mode ~~~~           & 
~~~~ mean~~~~  
~~~~ & variance~~~~  \\ 
\hline\hline   
\multicolumn{6}{|c|} 
{$\pi(\theta_k)=1, \quad \pi(\theta_k|n_k)=\Bc(n_k,N-n_k)$} 
\\ \hline
$n_k  $ & 
$N-n_k$ & 
$N$     & 
$\frac{n_k-1}{N-2}$ & 
$\frac{n_k}{N}$ & 
$\frac{n_k(N-n_k)}{N^2(N-1)}$ \\   
$n_k>0  $ & 
$n_k<N$ & 
$N>0$     & 
$\stack{n_k>0,}{N>2}$ & 
$\stack{n_k>0,}{N>0}$ & 
$\stack{n_k>0,}{\stack{n_k<N,}{N>1}}$ \\ \hline  
\multicolumn{6}{|c|} 
{$\pi(\theta_k)=\theta_k^{-1}(1-\theta)^{-1}, \quad \pi(\theta_k|n_k)=\Bc(n_k+1,N-n_k)$} 
\\ \hline
$n_k+1$ & 
$N-n_k$ & 
$N+1$   & 
$\frac{n_k}{N-1}$ & 
$\frac{n_k+1}{N+1}$ & 
$\frac{n_k+1)(N-n_k)}{(N+1)^2(N+2)}$ \\ 
$n_k>0  $ & 
$n_k<N  $ & 
$N>0$     & 
$\stack{n_k\ge 0,}{N>1}$ & 
$\stack{n_k\ge 0,}{N\ge 0}$ & 
$\stack{n_k\ge 0,}{\stack{n_k\le N,}{N\ge 0}}$ \\ \hline  
\multicolumn{6}{|c|} 
{$\pi(\theta_k)=\theta_k^{a-1}(1-\theta)^{b-1}, \quad \pi(\theta_k|n_k)=\Bc(n_k+a,N+b-n_k)$} 
\\ \hline
$n_k+a $  & 
$N-n_k+b$  & 
$N+a+b$& 
$\frac{n_k+a-1}{N+a+b-2}$& 
$\frac{n_k+a}{N+a+b}$ & 
$\frac{(n_k+a)(N-n_k+b)}{(N+a+b)^2(N+a+b+1)}$ \\  
$n_k\ge 0  $ & 
$n_k\le N  $ & 
$N\ge 0$     & 
$\stack{n_k\ge 0,}{N>0}$  & 
$\stack{n_k\ge 0,}{N\ge 0}$ & 
$\stack{n_k\ge 0,}{\stack{n_k\le N,}{N\ge 0}}$ \\ 
\hline\hline
\etabu
\caption{Different \apost laws corresponding to different choices of 
\aprio laws}
\end{table}

\bigskip 
Note also that, when we have the expression of the \apost law 
$\pi(\thetab|\nb)$, we may define other estimators than the MAP or 
the posterior mean (PM). 
We may also answer the questions of type $P(a<\theta_k<b)$. 
 
Note however that, all these computed numbers depend on the data 
and our prior knowledge we included. For any other data set we obtain 
other numbers. One may want to study the sensitivity of the solution 
to a kind of variability of data. This can be done by Monte Carlo 
simulations or by repeating the experience (but very often this may not 
be possible). 

Also, in general the sample size or, more precisely, the 
contrast between the sample size and the number of parameters, 
is a crucial parameter. One may want to know the convergence of 
the solution to the hypothetical case where the sample size 
goes to infinity. 

Now, let's see if we can answer some of the questions at the end of the 
last section. 
\bit
\item What is the probability of seeing face $k$ up based on the data 
set 1 or the data set 2? 
\\ 
For each data set, we can compute, for example, the following quantities: 
\bit
\item The most probable values $\theta_k^{MAP}$ of $\theta_k$;  

\item The mean values $\theta_k^{MP}$ of $\theta_k$;  

\item The variance values $v_k$ of $\theta_k$; 

\item The lower values $a_k$ and upper values $b_k$ for which 
the probabilities $P(a_k<\theta_k<b_k)=0.9$.   
\eit

\item Do these two data sets come from the same die? 
\\ 
We can try to answer this question by comparing the probability laws 
$\pi_1(\theta_k|\xb_1)$, $\pi_2(\theta_k|\xb_2)$ and  
$\pi(\theta_k|\xb_1,\xb_2)$. But how to do this comparison? 
We may try to compute the relative entropy 
\beq
KL(\pi_1 \pi_2; \pi)=\int \pi_1(\theta_k|\xb_1)\, \pi_2(\theta_k|\xb_2) 
\ln \frac
{\pi_1(\theta_k|\xb_1) \, \pi_2(\theta_k|\xb_2)}
{\pi(\theta_k|\xb_1,\xb_2)} \d{\theta_k}.
\eeq
If this value is near to zero, this means that the two data sets comes 
from different dice. 

\item Is this die loaded? 
\\ 
We can answer this question by computing the probabilities 
of two hypotheses $H_1=(\theta_1=\theta_2=\ldots=\theta_K)$ and 
$H_0=(\theta_k\not=\theta_l)$, \ie $P(H_1|\xb)$ and $P(H_0|\xb)$. 
\beqn
P(H_1|\xb)&=& \prod_k \int \d{\theta} \pi(\theta_k=\theta|\xb_1)  \\ 
P(H_0|\xb)&=& \int\d{\theta_1}\ldots\int\d{\theta_K} 
\prod_k \pi(\theta_k|\xb_1).
\eeqn

\item If I throw this die $N'=100$ times again, what will be the number of 
times I will see the face $k$ up? 
\\ 
To answer this question, there are two methods: \\ 
i) Use the data set $\xb=\{\nb,N,K\}$ to compute $\pi(\thetab|\xb)$ 
and estimate $\thetabh$ 
by one of the previous methods (MAP, PM, ...) and then compute 
$P(\nb'|\thetabh, N'=100, K)$. \\ 
ii) Try to find the expression of $P(\nb'|\nb,N,K,N')$ by following 
\beqnx
P(\nb|\thetab,N,K)
 &=& \prod_k C_N^{n_k}\, \theta_k^{n_k}\, (1-\theta_k)^{N-n_k},\\ 
P(\nb'|\thetab,N',K)
 &=& \prod_k C_{N'}^{n'_k}\, \theta_k^{n'_k}\, (1-\theta_k)^{N'-n'_k},\\ 
P(\nb,\nb'|\thetab,N,K,N')
 &=& \prod_k C_{N+N'}^{n_k+n'_k}\, \theta_k^{n_k+n'_k}\, (1-\theta_k)^{N+N'-n_k-n'_k},\\ 
P(\nb'|\nb,\thetab,N,K,N')
 &=& P(\nb,\nb'|\thetab,N,K,N') / P(\nb'|\thetab,N',K)
\eeqnx
and then integrate out $\thetab$ to obtain $P(\nb'|\nb,N,K,N')$. 
\eit

\section{Problem 3} 

Now, consider the case where, the observer has given to us only a subset 
$(n_1,\ldots,n_{K'})$ of the whole data $\nb=(n_1,\ldots,n_K)$ with 
$K'<K$. (He just has forgotten to count and report the numbers 
$\acc{n_k, k=K'+1,\ldots,K}$, but he is sure that the die has $K$ 
faces.  
In this case we can only obtain an expression for the likelihood function 
if we know the total number of the observations 
$N=\sum_{k=1}^K n_k \ge N'=\sum_{k=1}^{K'} n_k$ which is
\beq
P(\xb|\thetab) \propto 
\prod_{k=1}^{K'} \theta_k^{n_k}\, (1-\theta_k)^{N-n_k}.
\eeq

Note that this likelihood expression does not depend on the parameters \\ 
$\acc{\theta_k, k=K'+1,\ldots,K}$. 
Thus, the maximum likelihood (ML) estimation approach is unable to propose 
any values for them, while the Bayesian approach and in particular the MAP 
estimation can propose a solution which depends on the choice of \aprio. 
For example, with a uniform prior, we have:

\beq
\theta_k=\left\{\barr{ll} 
\frac{n_k}{N}  & k=1,\ldots,K' \\ 
\frac{(N-N')}{(K-K')N} & k=K'+1,\ldots,K
\earr\right. 
\label{thetak_Pb3}
\eeq
where the first row is common with ML and the second row is due to the uniform 
prior and the normalization. 

It is important to note that, while in the two previous cases, the prior 
law $\pi(\thetab)$ has a less important role, here the classical ML approach 
cannot give any answer the problem and the role of prior information is crucial. 
\section{Problem 4}  

Another interesting case is the one where we do not know the number of 
states (faces of the die). For example, we have observed the following data:
\\~\\ 
$\xb=[
4,2,2,2,1,*,4,*,1,4,3,3,*,*,4,*,*,4,4,1,1,2,1,*,4,2,4,2,3,2,2,*,2,2,1,\\
*,*,*,*,3,*,4,2,2,4,4,4,3,*,*,4,*,2,*,3,*,2,*,1,3,3,4,3,1,3,3,*,3,3,2,\\ *,*,3,4,4,3,3,3,4,1,2,4,4,*,4,*,*,*,*,*,*,*,1,1,4,1,*,2,1,*]$ 
\\~\\ 
where $*$ may mean  
\emph{anything else greater than 4} or 
\emph{do not know}. 
Note that these two cases are different. 
In the following, we first consider the first case which is, in fact, 
very close to the Problem~3 in the previous section, because we 
know exactly the 
$n_k$ for $k=1,\ldots,K'$ but we do not know other $n_k$, $k>K'$ 
nor the the true value of $K>K'$ itself.   
However, $N$ is given. We can only give an expression for the likelihood 
if we fix the value of $K$. Then, we can consider $K=5,6,7,...$ and for 
each case compute the results using~(\ref{thetak_Pb3}):\\~\\  
For $K=5$ we obtain: $\thetab=[0.1300,0.1700,0.1700,0.2100,0.03200]$ \\ 
For $K=6$ we obtain: $\thetab=[0.1300,0.1700,0.1700,0.2100,0.01600,0.01600]$ \\ 
For $K=7$ we obtain: $\thetab=[0.1300,0.1700,0.1700,0.2100,0.01067,,0.01067,0.01067]$ \\ 
and so on.
\\~\\                                                   

A difficult question remains: How to fix $K$? 
We may try to compare $\pi(\thetab|\xb,K)$ for different values of $K$ 
through their entropies. We may also choose a prior for it and compute 
$\pi(\thetab,K|\xb)$ or still 
integrate out $\thetab$ to obtain $\pi(K|\xb)$ from which we can estimate 
$K$. 

The case where, the $*$ in the data means \emph{do not know} is more 
complex. 
If at least we know $K$, then it may still be possible to write the 
expression of the likelihood. 
Let's note the true values of $n_k$ by $N\nu_k$. Then, we know that 
$N\nu_k=\in[n_k, n_k+n_*], \; k=1,\ldots,K'$ and 
$N\nu_k=\in[0, n_*], \; k=K',\ldots,K$ with 
$n_*=N-\sum_{k=1}^{K'} n_k$. Then, we may write

\beqnx
P(\xb|\thetab,\nub,K,K') &=& 
\prod_{k=1}^{K'} C_{N-n_*}^{n_k}\,\theta_k^{n_k}\,(1-\theta_k)^{N-n_*-n_k} 
\prod_{k=1}^{K}  C_{n_*}^{\nu_k}\,\theta_k^{N\nu_k}\,(1-\theta_k)^{n_*-N\nu_k}
\eeqnx
or
\beqnx
P(\xb|\thetab,\nub,K) &=& 
\prod_{k=1}^{K} C_{N}^{N\nu_k}\, \theta_k^{N\nu_k}\, (1-\theta_k)^{n_*-N\nu_k}. 
\eeqnx
We can then try to integrate out $\thetab$ from this expression to obtain 
$P(\xb|\nub,K)$ or integrate out $\nub$ to obtain $P(\xb|\thetab,K)$. 
But, what to do if we do not know $K$? Can we also integrate out $K$ by 
summing over all values of $K$? 

Another question that may arise in this problem and the previous ones, 
is to estimate the frequencies $\nu_k=n_k/N$ which is not exactly 
the same question of estimating $\theta_k$. 
In the following, we consider this problem. 

First consider the case of complete data $\{\nb,N,K\}$ of 
problems 1 and 2. 
We may note that, if we assume that the die is fair, the knowledge 
of the past experience ($\{\nb,N,K\}$) does not change anything on the 
results  of the future experience. But, if we do not know if the die 
is loaded, then from the past experience, we can estimate $\thetab$ 
and use it to compute the probability of observing any event. 

The situation becomes more complex if we do not know $K$ or $N$ or 
if some data are missing as is the case in problems 3 or 4, or more 
generally the cases where we cannot write easily the exact expression 
of the likelihood.  

Consider the incomplete data problem 4 where 
we know $N$, $n_k$ and $n_{*}$, but we do not know $K$ and 
assume that the $*$ are \aprio distributed uniformly between 
$1$ and $K$ (or between $K'$ and $K$) and compute the numbers 
$d_k=(n_k+n_*/K)/N$ (or $d_k=(n_k+n_*/(K-K'))/N$). 
We can then say that these computed $d_k$ are good approximations 
to the true unobserved $\nu_k$. 
The question is how to model this approximation.  
Two models can then be used:
\bit 
\item[i)] Assume $d_k$ as the mean values of the unknown frequencies 
$\nu_k$ 
\beq
d_k=\esp{\nu_k} = \int \nu_k p(\nu_k) \d{\nu}_k
\eeq
or  
\item[ii)] Assume each $d_k$ to be the sum of the true  $\nu_k$ and a 
random error $\epsilon_k$:
\beq
d_k = \nu_k + \epsilon_k
\eeq
where $\epsilon_k$ is assumed to be centered with unknown pdf. 
In both cases, we are interested in finding  $p(\nu_k|d_k)$ or 
$p(\nub|\db)$. 
\eit 
But, before going further, it is important to note that, 
in the following, we are not going to analyze the original 
data $\xb$ but the \emph{pre-processed} data $\db$. 
We changed the problem to a new one: Given $\db$ can we assign or 
compute $p(\nub|\db)$. \\ 
Two approaches can then be used.  

\bigskip\noindent{\bf Information Theory or Maximum Entropy approach:}
\\[6pt] 
This approach is based on the first equation between $d_k$ and $\nu_k$.  
It is obvious that, there are an infinite number of possible solutions to 
this equation. Let us denote by $\Pc$ this ensemble:
\beq
\Pc=\{p~:~ \esp{\nu_k} = \int \nu_k p(\nu_k) \d{\nu}_k = d_k\}.
\eeq
The Maximum Entropy principle chooses the one $p^{ME}(\nu_k)$ with the 
highest entropy   
\beq
p^{ME}(\nu_k)=\argmax{p\in\Pc}{H(p)} 
\eeq
where
\beq
H(p)= - \int p(x) \ln p(x) \d{x},  
\eeq
or, more generally, if we assume a reference (prior?) distribution 
$q(\nu_k)$, the one $p^{MKL}(\nu_k)$ which has minimum Cross Entropy 
or Kullback-Leibler (KL) divergence 
\cite{Shore80,Kullback51,Kullback59},  
of $p$ with respect to to $q$:  
\beq
p^{MKL}(\nu_k)=\argmin{p\in\Pc}{KL(p,q)} 
\eeq
where 
\beq
KL(p,q)=\int p(x) \ln (p(x)/q(x)) \d{x}.  
\eeq
We note that when $q$ is uniform $KL(p,q)=-H(p)$ and thus $p^{MKL}(\nu_k)=p^{ME}(\nu_k)$. 

The unique solution, if exists, is given by
\beq
p^{MKL}(\nu_k)=\frac{1}{Z(\lambda_k)} q(\nu_k) 
\expf{-\lambda_k \nu_k} 
\eeq
where 
\beq
Z(\lambda_k)=\int q(\nu_k) \expf{-\lambda_k \nu_k} \d{\nu_k}, 
\eeq
and it can be shown that $\lambda_k$ is the 
solution of the equation
\beq
- \partial \ln Z(\lambda_k) / \partial \lambda_k=d_k 
\label{LambdaSolution}
\eeq
which can be computed numerically. It is evident that the expressions of $p^{MKL}(\nu_k)$, $Z(\lambda_k)$, and consequently any numerical values 
for the estimate   
\beq
\nu_k^{MKL}=\esp{\nu_k} = \int \nu_k p^{MKL}(\nu_k) \d{\nu}_k 
\label{MKLSolution}
\eeq
depend on the choice of $q$. 

As a matter of algorithmic and computation of $\lambdabh$ (solution of the 
equation~(\ref{LambdaSolution})) 
and $\nubh$ defined in~(\ref{MKLSolution}), it is interesting to know 
that they can be computed through: 
\beq
\left\{\barr{ll}
\lambdabh&=\argmin{\lambdab}{D(\lambdab)=\ln Z(\lambdab)+\lambdab^t\db}, \\ 
\nubh    &=\argmin{\nub\in\Cc}{H(\nub,\nub^{(0)})}
\earr\right.
\eeq
where $D(\lambdab)$ is called the \emph{dual criterion} 
and $H(\nub,\nub^{(0)})$ 
is called the \emph{primal criterion} and where 
$\nu_k^{(0)}=\espx{q}{\nu_k}=\int \nu_k q(\nu_k) \d{\nu_k}$. 

The expressions of dual and primal criteria also depends on 
the expression of $q$. 
For example, when $q$ is uniform on $\Cc$, $p$ is exponential 
we have 
\[
Z(\lambdab)=\prod_k (1/\lambda_k),\;  
\ln Z(\lambdab)=-\sum_k \ln \lambda_k,\; 
D(\lambdab)=-\sum_k \ln \lambda_k +\sum_k \lambda_k d_k 
\]
and 
\[
P(\nub,\nub^{(0)})=-\sum_k \ln (\nu_k/\nu_k^{(0)}) 
+\sum_k (\nu_k-\nu_k^{(0)}). 
\] 
For other choices of $q$ and more details on these relations 
refer to 
\cite{Rockafellar70,Borwein91a,Djafari91a,Rockafellar93,LeBesnerais93a,Djafari94,Bercher95a,Djafari96g,LeBesnerais99,Djafari99b}.

\bigskip\noindent{\bf Bayesian approach:}\\[6pt]  
The Bayesian approach is based on the second equation, \ie, 
$d_k=\nu_k+\epsilon_k$ and we have to find an expression for 
the likelihood $\Lc(\nub)=P(\db|\nub)$ and assign a prior 
$q(\nu_k)$ or $q(\nub)$. When this done we can give an expression 
for the posterior $\pi^{B}(\nub|\db)$. 
Note that, in both cases, we have to choose $q(\nub)$. 
The first step, which is to find an expression for $\Lc(\nub)=P(\db|\nub)$,  
is not easy. Here are a few approaches:

\bigskip\noindent\emph{Assuming $\thetab=\nub$:} \\[6pt] 
The first approach consists in assuming $\thetab=\nub$. 
Then, if we are also given $N$, the problem becomes equivalent to the 
Problem~2 and we have:
\beq
P(\db|\nub,N)=\prod_k C_N^{N d_k} \nu_k^{N d_k} (1-\nu_k)^{N(1-d_k)}. 
\eeq
Then, again choosing a uniform prior 
$q(\nub)=\frac{1}{Z_0}\delta(1-\sum_k \nu_k)$, we obtain 
\beq
\pi(\nu_k|\db,N)=\Bc\left(N d_k-1, N(1-d_k)-1\right)
\eeq
and then we have 
\beq
\esp{\nu_k|\db,N}=\frac{N d_k-1}{N-2}.
\eeq
We see that $\esp{\nu_k|\db,N}\mapsto d_k$ when $N$ goes to infinity. 

But, if we do not know $N$, we can try to integrate out $N$. 
Can we do it easily? I did not go further in this direction. 

\bigskip\noindent\emph{Frequentist point of view:} \\[6pt]  
Here, we assume \aprio that the die is fair and try to obtain 
an expression for the likelihood $\Lc(\db|\nub,N)$ using the following 
arguments: \\   
Given $N$ and $K$ and assuming that each through of the die 
is independent of all others, 
we may argue on the number of possible outcomes resulting 
to a particular data set using the multinomial coefficient 
\beq
W(\nb,N,K)=\frac{N!}{n_1!\ldots n_k!}=\frac{N!}{\prod_{k=1}^K (n_k!)}.
\eeq
$W(\nb,N,K)$ is the number of possible outcomes $\xb$ such that the 
face $k$ appears $n_k$ times between the total possible outcomes 
which is $K^N$. 
Thus, we may assign 
\beq
P(\nb|N,K)=W(\nb,N,K)/(K^N)=\frac{N!}{(K^N) \prod_{k=1}^K (n_k!)}.
\eeq

It is known that, using the Stirling approximation
\footnote{Stirling (1692-1770) showed that 
$x_n=\frac{n! e^n}{n^{n+1/2}}$ converges to $\sqrt{2\pi n}$ when $n$ 
goes to $\infty$. This means that, for large $n$ we get the 
approximation $\ln(n!)=\frac{1}{2}\ln (2\pi n)+n\ln n$. 
However, even if this is usually called Stirling's formula, in fact, 
it may have been known earlier to Abraham de Moivre (see  {\tt http://www-gap.dcs.st-and.ac.uk/\~history/Mathematicians/De\_Moivre.html}).} 
the expression of this probability, when $N$ is large, 
converges to 
\beq
\lim_{N\mapsto\infty} \ln P(\nb|N,K)=H(\nub)=-\sum_{k=1}^K \nu_k \ln \nu_k 
\eeq
where $\nu_k=\lim_{N\mapsto\infty} \frac{n_k}{N}$. 

This explanation and this approximation have also been used to justify 
the choice of an expression for entropy $H(\nub)=-\sum_k \nu_k \ln \nu_k$ 
and a prior law for $nb$ which is $\pi(\nb)\propto\expf{\alpha H(\nu)}$, 
so that, given a set of constraints on $\nu_k$, finding the 
most probable (sampling argument or maximum likelihood approach) 
value of $\nb$ subject to those constraints  
become equivalent to maximizing $H(\nub)$ subject to those constraints:  
\beq
\nbh_k=\argmax{\nb}{\ln \pi(\nb)}=\argmax{\nb}{H(\nb)}.
\eeq

But, we do not know either $N$ or $K$. We may however try to use these expressions 
to find approximations to the likelihood function we need. 
First, we may assign 
\beq
P(\db|\nub,N,K)=P(N\db|N,K)(1-P(N\nub|N,K)), 
\eeq
and replacing for $P(N\db|N,K)$ and $P(N\nub|N,K)$ and using again the 
Stirling formula we may find an expression which may be independent 
of $N$. 

 
\bigskip\noindent\emph{Integration of nuisance parameter $\thetab$:}\\[6pt] 
Again here, we start by assuming $N$ known. Then, we know the expressions 
of $P(\db|\thetab,N)$ and $P(\nub|\thetab,N)$:
\beq
P(\db|\thetab,N) = 
\prod_k C_N^{N d_k}\, \theta_k^{N d_k}\, (1-\theta_k)^{N(1-d_k)}
\eeq
and 
\beq
P(\nub|\thetab,N) = 
\prod_k C_N^{N \nu_k}\, \theta_k^{N \nu_k}\, (1-\theta_k)^{N(1-\nu_k)}.
\eeq
Then we can write 
\beqn
P(\db|\nub,\thetab,N) 
&=& \left(1-P(\nub|\thetab,N)\right) \; P(\db|\thetab,N) \nonumber\\ 
&=& \left(
1-\prod_k C_N^{N \nu_k}\, \theta_k^{N \nu_k}\, (1-\theta_k)^{N(1-\nu_k)}
\right) \nonumber\\ 
&& \times \prod_k C_N^{N d_k}\, \theta_k^{N d_k}\, (1-\theta_k)^{N(1-d_k)}.
\eeqn
Then, we have to integrate out $\thetab$ to obtain the likelihood 
$\Lc(\nub)=P(\db|\nub,N)$. 
Can we obtain simple expressions? Can we integrate out $N$ too? 
I did not go farther in this direction. 

\bigskip\noindent\emph{Ad hoc empirical approach:} \\[6pt] 
Another approach is to assign the two pdfs 
$p(\epsilon)=p(d_k-\nu_k)$ 
and the prior $q(\nu_k)$ from which we can compute 
\beq
\pi^{B}(\nu_k|\d_k)=p(d_k-\nu_k)\, q(\nu_k)~/~m(d_k).
\eeq 
Here too, the expression of the posterior pdf $\pi(\nu_k|\d_k)$ and 
thus any inference about $\nu_k$ depends on the choice of 
$p(\epsilon)$ and $q(\nu_k)$. 

A question may arise here: 
\\ 
Can we first fix $q(\nu_k)$ and compute $p^{MKL}(\nu_k)$ and 
use it again as a prior in this Bayesian approach? 
\\ 
The answer is "No", because $p^{MKL}(\nu_k)$ is in fact 
$p^{MKL}(\nu_k|d_k)$ 
and doing so, we have used two times the same data $d_k$. 

Another question is how to compare and how to use $p^{MKL}(\nu_k|d_k)$ 
and $\pi^{B}(\nu_k|\d_k)$? 
\\ 
My answer is that $\pi^{B}$ contains more 
information than that of $p^{MKL}$, because to obtain $\pi^{B}$, 
we combined information about both $\epsilon_k$ through $p(\epsilon)$ 
and $\nu_k$ through $q(\nu_k)$ while 
to obtain $p^{MKL}$ we used only $q(\nu_k)$. 
Indeed, it seems that 
the only consistent point estimator of $\nu_k$ from $p^{MKL}$ is 
its posterior mean, while, 
there is not any such restriction on $\pi^{B}$. 

\section{Problem 5}  

An important case is the one where we have only given the mean value of the 
face numbers $\sum_k k \, \theta_k=d_0$ or the more general case of the mean 
value of the numbers written on the faces $\sum_k g_k \, \theta_k=d$ without 
any other knowledge and, in particular, without knowing $N$. We need however 
to know $K$. 

Remember also that $\esp{X}=\sum_k k \, \theta_k$ and 
$\esp{G}=\sum_k g_k \, \theta_k$ are not the same. 
They become equivalent if $g_k=k$. 

Thus, we consider the case:
\beq
\sum_k g_k \, \theta_k=d
\label{c1}
\eeq
and we assume to know the number of states $K$. 
The objective is to find $\theta_k$.  

\bigskip\noindent{\bf MaxEnt solution:}\\[6pt]  
The classical answer this problem is MaxEnt which can be described as 
follows: 
\\ 
It is obvious that, there are infinite number of possible solutions to 
the equation~(\ref{c1}). 
The Maximum Entropy principle chooses the one with the highest 
entropy   
\beq
H(\thetab)=-\sum_k \theta_k \, \ln \theta_k.
\eeq
The solution has the form 
\beq
\theta_k(\lambda)=\frac{1}{Z(\lambda)}\expf{-\lambda \, g_k}
        =\expf{-(\ln Z(\lambda)+\lambda \, g_k)}, 
\eeq
where 
\beq
Z(\lambda)=\sum_k \expf{-\lambda \, g_k}, 
\eeq
and $\lambda$ is the solution of the following equation
\beq
- \partial \ln Z(\lambda) / \partial \lambda=d 
\eeq
which can be computed numerically. 

It is also easy to show that the maximum value of the entropy is 
\beq
H_{max}(\thetab)=-\sum_k \theta_k \, \ln \theta_k 
=\ln Z(\lambda) + \lambda d 
=\max_{\lambda} \ln \theta_k(\lambda)
\eeq
which can also be written 
\beq
\max_{\lambda} \thetab(\lambda)= \expf{H_{max}(\thetab)}.
\eeq

\bigskip\noindent{\bf Bayesian solution:}\\[6pt]  
If we knew $N$, we could write the expression of the likelihood 
$P(D=d|\thetab,N)$ with $d=\sum_k g_k n_k$ and $\sum_k n_k=N$: 
\beq
P(D=d|\thetab,N)=\prod_{n_k=0}^{N} P(\nb|\thetab) 
\delta(N-\sum_k n_k)\delta(d-\sum_k g_k n_k).
\eeq
We can also try to integrate out $N$:
\beq
P(D=d|\thetab)=\sum_{N=0}^{\infty} 
\prod_{n_k=0}^{N} P(\nb|\thetab) \delta(N-\sum_k n_k)
\delta(d-\sum_k g_k n_k).
\eeq
These computations seem to me intractable. In the following, I propose 
another approach:

The main idea here is that, we may account for uncertainty of 
this data (in particular, because we do not know the value of $N$) 
by assuming 
\beq
p(d|\thetab)=\Nc\pth{d-\sum_k g_k \, \theta_k, \sigma^2}, 
\eeq
and by arguing on the additivity and positivity of $\thetab$ we choose 
\beq
\pi(\thetab)=\expf{-H(\thetab)}.
\eeq
Then, the posterior is
\beq
\pi(\thetab|d)=
\expf{-\frac{1}{2\sigma^2}(d-\sum_k g_k \, \theta_k)^2-H(\thetab)}, 
\label{posterior1}
\eeq
and the MAP solution is 
\beq
\thetabh=\argmin{\thetab}{(d-\sum_k g_k \, \theta_k)^2-\alpha H(\thetab)}
\label{MAP1}
\eeq
with $\alpha=2\sigma^2$. 

Now, if we choose $H(\thetab)=\sum_k \theta_k \, \ln\theta_k$ the 
numerical results obtained by this approach and those obtained by using 
the MaxEnt solution become almost identical. 
However, if we can fix the value of $\alpha$, we have access to 
the $\pi(\thetab|d)$ which contains more information than only one point 
estimator. 

\bigskip\noindent{\bf Combined data fusion solution:}\\[6pt]  
Assume now that, not only we have the data $\xb$ or $\nb$, but also $d$ 
from previous section. How to combine them? 
Here is my solution. 

Follow the Bayesian approach of the sections 1 or 2 to write down the 
expression of the \apost law 
\beq
\ln \pi(\thetab|\nb)=
\sum_k \cro{ {n_k}\ln \theta_k + (N-n_k)\ln(1-\theta_k)} +\ln \pi(\thetab)+c
\eeq
and use the expression of $\pi(\thetab|d)$ in equation~(\ref{posterior1}) 
as the prior $\pi(\thetab)$ here. 

\section{Problem 6}

Assume now that, our observer has repeated the experience $L$ times, 
and before each experience, he has changed the numbers written on each face. 
For example, the first time, he has written $g_k=k$ and for the second experience $g_k=k^2$. This is also equivalent to the experiment of using $L$ 
similar dice with different colors and different labeling on each faces simultaneously. 
Then, he computed the numbers $n_{kl}$. 

But, assume now that, finally, he gives us only the 
mean values $\bar{n}_l=(1/N)\sum_k n_{kl}$ or $d_l=(1/N)\sum_k g_{kl}$.  
The problem is similar to the previous case, but here we have $L$ 
data: 
\beq
 \sum_k g_{kl} \, \theta_k=d_l, \quad l=1,\ldots,L, 
\eeq
which can be written $\Gb \thetab=\db$ where $\Gb$ is the matrix with 
elements $g_{kl}$. 
Thus, we have a linear system of equations with $K$ unknowns and $L$ data. 
Note that here we know exactly the values $g_{kl}$. 

If the experimenter has made good choices for $g_{kl}$ and if $L=K$, 
then we may only try to solve that system of equations and obtain an 
exact solution to the problem. 
But, what if $L<K$ or if the experimenter has not made a good 
choice for $g_{kl}$, for example, if he has naively written $g_{kl}=kl$. 
In both cases, the system of equations has an infinite number of 
solutions.  

\bigskip\noindent{\bf MaxEnt solution:}\\[6pt]  
The MaxEnt approach is again straightforward and  
the solution has the form 
\beq
\theta_k=\frac{1}{Z(\lambdab)}\expf{-\sum_l \lambda_l \, g_{kl}}
        =\expf{-(\ln Z(\lambdab)+\sum_l \lambda_l \, g_{kl})}, 
\eeq
where 
\beq
Z(\lambdab)=\sum_k \expf{-\sum_l \lambda_l \, g_{kl}}, 
\eeq
and $\lambdab=[\lambda_1,\ldots,\lambda_L]$ is the 
solution of the following equation
\beq
- \partial \ln Z(\lambdab) / \partial \lambda_l=d_l 
\eeq
which can be computed numerically. 
It is also easy to show that the maximum value of the entropy is 
\beq
H_{max}(\thetab)=-\sum_k \theta_k \, \ln \theta_k 
=\ln Z(\lambda) + \lambdab^t\db
=\max_{\lambda} \ln \thetab(\lambdab)
\eeq
which can also be written 
\beq
\max_{\lambdab} \thetab(\lambdab)= \expf{H_{max}(\thetab)}.
\eeq

\bigskip\noindent{\bf Bayesian solution:}\\[6pt]  
Following the steps of the section 5, we have
\beq
p(\db|\thetab)=\Nc\pth{\db-\Gb\thetab, \sigma^2}
\eeq
and by arguing on the additivity and positivity of $\thetab$ we choose 
\beq
\pi(\thetab)=\expf{-H(\thetab)}.
\eeq
Then, the posterior is
\beq
\pi(\thetab|d)=
\expf{-\frac{1}{2\sigma^2}\|\db-\Gb\thetab\|^2-H(\thetab)}
\label{posterior2}
\eeq
and the MAP solution is 
\beq
\thetabh=\argmin{\thetab}{\|\db-\Gb\thetab\|^2-\alpha H(\thetab)}
\label{MAP2}
\eeq
with $\alpha=2\sigma^2$. 

\bigskip\noindent{\bf Combined data fusion solution:}\\[6pt]  
Assume now that, not only we have the data $\xb$ or $\nb$, 
but also $\db$ from the previous section. How to combine them. 
Here again we can follow the Bayesian approach of the sections 1 
or 2 to write down the expression of the \apost law 
\beq
\ln \pi(\thetab|\nb)=
\sum_k \cro{ {n_k}\ln \theta_k +(N-n_k)\ln(1-\theta_k)}+\ln \pi(\thetab)+c
\eeq
and use the expression of $\pi(\thetab|\db)$ in equation~(\ref{posterior2}) 
as the prior $\pi(\thetab)$ here. 

\section{Problem 7} 

Consider the same previous experiment, but this time, the experimenter 
is sure that all dice were absolutely identical and unloaded, but he 
has forgotten to note the numbers he has written on the dice faces. 

However, he has also noted the mean values $(1/L)\sum_l g_{kl}=d_k$. 
Can we be of any help for him to find them? 

Thus, this time, $\theta_k=1/K, k=1,\ldots,K$ and we have 
\beq
 \sum_k g_{kl} \, \theta_k=(1/K) \sum_k g_{kl}=d_l, \quad l=1,\ldots,L, 
\eeq
and also $(1/L)\sum_l g_{kl}=d_k, \quad k=1,\ldots,K$. 

The problem becomes an interesting one, we want to compute the elements 
of a matrix from its row and column sums. 
This mathematical problem arises in many other applications such as 
computed tomography where we want to recover the pixel values of an image 
from its horizontal and vertical projections.  

Except the case of $K=L=2$, we have 
always less data than unknowns and the problem has an infinite number of solutions. Even in the case $K=L=2$ where the number of unknowns and data 
are equal, the problem is still under-determined and 
has infinite number of solutions. 

We need to question our experimenter to see if he can remember 
of any other information about those numbers (prior information or 
constraints?) which can be helpful to give reasonable answers 
about this question. 

To go further in details of this problem, let's change slightly the 
notation. We want to estimate the elements $g_{kl}$ of a 
$(K\times L)$ matrix $\Gb$ from its row sums 
$r_k=\sum_l g_{kl}$ and its column sums $c_l=\sum_k g_{kl}$. 

We may also note 
$\rb=[r_1,\ldots,r_K]$, $\cb=[c_1,\ldots,c_L]$, $\db=[\rb;\cb]$ and  
$\gb$ a vector containing all the elements of the matrix $\Gb$ 
concatenated column by column. 
Then, it is easy to see that we can also write $\cb=\Ab_1\gb$, 
~~$\rb=\Ab_2\gb$ and thus $\db=\Ab\gb$ where 
$\Ab_1$, $\Ab_2$ and $\Ab$ are, respectively, a $(K \times KL)$, 
a $(L \times KL)$ and a $((K+L) \times KL)$ matrix with 
$\Ab=\pmatrix{\Ab_2\\ \Ab_2}$ and whose elements are composed of 
zeros and ones. 

Now, we consider two sets of answers of our experimenter: 
those who put deterministic constraints on $g_{kl}$ and those who 
put probabilistic constraints. 
 
\bigskip\noindent{\bf Deterministic constraints:} 
\bit
\itema $g_{kl}=g_k$. Then, we have $r_k=L g_k$ and we have a unique 
solution $g_k=r_k/L$ subject to the condition that 
$\sum_k g_k=\frac{K}{L}\sum_k r_k=c_l, \quad l=1,\ldots,L$. 

\itema $g_{kl}=g_l$. Then, we have $c_l=K g_l$ and we have a unique 
solution $g_l=c_l/K$ subject to the condition that 
$\sum_l g_l=\frac{L}{K}\sum_l c_l=r_k, \quad k=1,\ldots,K$. 

\itema $g_{kl}=g_{1_k}\,g_{2_l}$. 
Then, we have $r_k=g_{1_k} \sum_l g_{2_l}$ and 
$c_l=g_{2_l} \sum_k g_{1_k}$ 
and we have $g_{1_k}\propto r_k$ and $g_{2_l}\propto c_l$. 
There still remains  two unknowns $\sum_l g_{2_l}$ and 
$\sum_k g_{1_k}$. However, if $g_{1_k}$ and 
$g_{2_l}$ are normalized, then we have a unique solution. 

\itema $g_{kl}$ are normalized as they represent a probability 
distribution:  
$\sum_k g_{kl}=\sum_l g_{kl}=\sum_k \sum_l g_{kl}=1$. 
This information is not enough to find a unique solution. 
That becomes true if $g_{kl}$ is separable as in the previous case. 

\itema $g_{kl}$ are normalized as they represent a probability 
distribution: 
$\sum_k g_{kl}=\sum_l g_{kl}=\sum_k \sum_l g_{kl}=1$ 
and and are distributed as uniformly as possible over the 
grid~$\{(k,l), k=1,\ldots,K, l=1,\ldots,L\}$.  

\smallskip 
This information may be enough to find a solution if it exists, 
by maximizing 
$H(\gb)=-\sum_j g_j\ln g_j$ subject to the data constraint 
$\Ab\gb=\db$ and 
the normalization constraint $\sum_j g_j=1$. 
Then the solution is given by 
$\gb=\frac{1}{Z(\lambdab)}\expf{\Ab^t\lambdab}$ 
where $\lambdab$ is the solution of 
$-\partial\ln Z(\lambdab)/\partial\lambda_j=d_j$ 
which can also be computed by 
$\lambdabh=\argmin{\lambdab}{D(\lambdab)=\ln Z(\lambdab)+\lambdab^t\db}$. 

\smallskip 
Unfortunately, there is not an explicit expression for this solution, 
but it is by construction positive 
$(g_j\propto\expf{[\Ab^t\lambdab]_j})$ 
and satisfies the data and normalization constraints for any correct 
data sets. 
Note also that this solution is not a linear function of the data.  

\smallskip 
There is only one question remaining: Is there any other 
criteria $H(\gb)$ which can give these satisfactions? 

\smallskip 
To give a partial answer to this question, we may say that any convex 
criterion can be used to find a unique solution. 
For example, $H(\gb)=\sum_j g_j^2=\|\gb\|^2$ which gives the minimum 
norm (generalized inverse) solution $\gb=\Ab^{+}\db$ which becomes $\gb=\Ab^t(\Ab\Ab^t)^{-1}\db$ 
if $\Ab\Ab^t$ was invertible. Note that this solution is a linear 
function of the data, but, this criterion does not guarantee the 
positivity of the solution. 

\smallskip 
Another example is $H(\gb)=\sum_j \ln g_j$ which gives the solution of 
the form $g_j=\frac{1}{[\Ab^t\lambdab]_j}$ but, this criterion does not 
guarantee neither the positivity or the boundedness of the solution. 
One can find other convex criteria (see next section). 
\eit

\bigskip\noindent{\bf Probabilistic constraints:} 
\bit
\itema We know that $\gb\in\Cc$ and that we generated $\gb$ according to  
a reference measure $q(\gb)$ over $\Cc$ such that $\espx{q}{\gb}=\gb_0$. 
Now, again, we can use the ME tool and search for $p(\gb)$ such that 
$\Ab\espx{p}{\gb}=\db$ and minimizes $KL(p,q)$. 
We know that the solution 
is $p(\gb)=\frac{1}{Z(\lambdab}q(\gb)\expf{\lambdab^t\Ab^t\gb}$ where 
$\lambdab$ is the solution of 
$-\partial\ln Z(\lambdab)/\partial\lambda_j=d_j$ which can also be 
computed by 
$\lambdabh=\argmin{\lambdab}{D(\lambdab)=\ln Z(\lambdab)+\lambdab^t\db}$ 
and finally, the solution  
$\gbh=\espx{p}{\gb}$ can be computed by 
$\gbh=\argmin{\Ab\gb=\db}{H(\gb,\gb_0}$. 
However, as we discussed it before, the expression of $H$ depends on 
the choice $q(\gb)$:  

\smallskip 
For $\Cc$ a closed set of real numbers and $q(\gb)$ Gaussian, we have \[
H(\gb,\gb^{(0)})=\|\gb-\gb_0\|^2.
\] 
For for $\Cc$ a closed set of real numbers and $q(\gb)$ a Lebesgue 
measure on $\Cc$, we have 
\[
H(\gb,\gb^{(0)})=-\sum_j \ln (g_j/g_{0_j})+(g_j-g_{0_j}), 
\] 
and, finally, 
\\ 
For $\Cc$ a closed set of integer numbers and $q(\gb)$ Poissonian, 
we have 
\[
H(\gb,\gb^{(0)})=KL(\gb,\gb_0)=\sum_j g_j \ln (g_j/g_{0_j})+(g_j-g_{0_j}).
\] 

\smallskip 
This discussion shows a relation between the classical ME approach of the 
last section and the ME in the mean as is presented here. Even if here, 
we have a tool to derive the expression of the needed convex criterion, 
still an arbitrary remains on the choice of $\Cc$ and the reference 
measure $q(\gb)$. 

\itema Each element $g_{kl}$ has been generated independently using a 
Gaussian random number generator: 
$g_{kl}\sim \Nc(k,\lambda)$. 

\itema Each element $g_{kl}$ has been generated independently using a 
Gaussian random number generator: 
$g_{kl}\sim \Nc(l,\lambda)$. 

\itema Two sets of numbers $g_{1_k}$ and $g_{2_l}$ have been generated 
using a Gaussian random number generator $g_{1_k}\sim \Nc(k,\lambda_1)$ 
and $g_{2_l}\sim \Nc(l,\lambda_2)$, then normalized and point-wise 
multiplied: $g_{kl}=g_{1_k}g_{2_l}$. 
 
\itema Each element $g_{kl}$ has been generated independently using a 
random number generator. We consider two interesting cases:  
$g_{kl}\sim \Nc(\mu,\lambda)$ and  
$g_{kl}\sim \Pc(\lambda)$. 

\itema The elements $g_{1l},g_{k1},g_{1L},g_{K1}$ have been generated  independently using a 
random number generator $\Nc(0,1)$, but others are generated by 
$g_{kl}\sim\Nc(\bar{g}_{kl},1)$ where $\bar{g}_{kl}=\frac{1}{4}[g_{k-1,l}+g_{k,l-1}+g_{k+1,l}+g_{k,l+1}]$. 
\eit

Let's consider only the case of independent Gaussian 
$g_{kl}\sim \Nc(\mu,\lambda)$ and $g_{kl}\sim \Pc(\lambda)$ where 
we may be able to do all the computations. 

\bigskip\noindent{\bf Gaussian case:}\\[6pt]  
We have: 
\beqnx
g_{kl}\sim \Nc(\mu,\lambda) \lra  p(g_{kl})=(\frac{1}{2\pi\lambda})^{\frac{1}{2}} 
 \expf{-\frac{1}{2\lambda}(g_{kl}-\mu)^2}.  
\eeqnx
Then, the column sums $c_l$ and rows sums $r_k$ are also Gaussian:
\beqnx
 r_k=\sum_l g_{kl}\sim\Nc(L\mu,L\lambda), \quad 
 c_l=\sum_k g_{kl}\sim\Nc(K\mu,K\lambda), 
\eeqnx
and thus:
\beqnx
 p(\rb)=
 (\frac{1}{2\pi L})^{\frac{K}{2}} 
 \expf{-\frac{1}{2L\lambda}\sum_k(r_k-L\mu)^2} 
\eeqnx
\beqnx
 p(\cb)=
 (\frac{1}{2\pi K})^{\frac{L}{2}} 
 \expf{-\frac{1}{2K\lambda}\sum_l(c_l-K\mu)^2}.
\eeqnx
Then, we can write the expression of the posterior law:
\beqnx
 p(g_{kl}|\rb,\cb,\lambda)&\propto& P(g_{kl},\rb,\cb|\lambda)
 =\expf{-\frac{1}{2\lambda}(g_{kl}-\mu)^2} \\ 
 &\times&
 \expf{-\frac{1}{2L\lambda}\sum_k(r_k-L\mu)^2}  
 \times  
 \expf{-\frac{1}{2K\lambda}\sum_l(c_l-K\mu)^2} \\
 & & \mbox{with~~~} r_k=\sum_l g_{kl}\mbox{~~and~~} c_l=\sum_k g_{kl}.
\eeqnx
It is then easily seen that 
\beqnx
 p(g_{kl}|\rb,\cb,\lambda)&\propto& \expf{-\frac{1}{2\lambda}J(g_{kl})} \\ 
 \mbox{with~~} J(g_{kl})&=& 
 (g_{kl}-\mu)^2
 +\frac{1}{K}\sum_k(\sum_l g_{kl}-L\mu)^2 
 +\frac{1}{L}\sum_l(\sum_k g_{kl}-K\mu)^2
\eeqnx
is Gaussian and we can easily compute its mean and variance. 
To obtain the mean values, we can compute the derivative of 
\beqnx
 J=(g_{kl}-\mu)^2
 +\frac{1}{K}\sum_k(r_k-L\mu)^2 
 +\frac{1}{L}\sum_l(c_l-K\mu)^2 
\eeqnx
which is
\beqnx
\partial J/\partial g_{kl} &=& 
2(g_{kl}-\mu)+\frac{2}{K} \sum_k(r_k-L\mu)
+\frac{2}{L} \sum_l(c_l-K\mu) 
\eeqnx
and equate it to zero to obtain
\beq
 g_{kl}=\frac{KL}{KL+2L+2K}\left(
 \mu+\frac{1}{K}\sum_k r_k +\frac{1}{L}\sum_l c_l\right)
\eeq
This result is interesting, because 
$\frac{1}{K}\sum_k r_k+\frac{1}{L}\sum_l c_l$ 
is what is called the back-projection in computed tomography. 

\bigskip 
We can generalize these results, if we work with the vectors 
$\gb$, $\rb=\Ab_1\gb$, $\cb=\Ab_2\gb$ and $\db=[\stack{\rb}{\cb}]=[\stack{\Ab_1}{\Ab_2}]\gb=\Ab\gb$. Then, we have:
\beqnx
 \gb\sim\Nc(\gb_0,\Rb_g), \quad 
 \db\sim\Nc(\Ab\gb_0,\Ab\Rb_g\Ab^t), \quad 
 \cro{\stack{\gb}{\db}} \sim\Nc\left(
 \cro{\stack{\gb_0}{\Ab\gb_0}},
 \cro{\stack{\Rb_g}{\Ab\Rb_g}~\stack{\Rb_g\Ab^t}{\Ab\Rb_g\Ab^t}}
 \right) 
\eeqnx
and thus
\beqnx
 \gb|\db\sim\Nc(\wh{\gb},\wh{\Rb}_g), \quad\mbox{with}\quad
 \left\{\barr{l} 
 \wh{\gb}=\gb_0+\Rb_g\Ab^t(\Ab\Rb_g\Ab^t)^{+}(\db-\Ab\gb_0) \\
 \wh{\Rb}_g=\Rb_g-\Rb_g\Ab^t(\Ab\Rb_g\Ab^t)^{+}\Ab\Rb_g
 \earr\right.,
\eeqnx
where $(\Ab\Rb_g\Ab^t)^{+}$ is the generalized inverse of $\Ab\Rb_g\Ab^t$. 

Note that when $\Ab\Rb_g\Ab^t$ is invertible, we have 
$\wh{\gb}=\Ab^{-1}\db$ and $\wh{\Rb}_g=0$. 

For the particular case of $\Rb_g=\lambda\Ib$ we have 
\beq
 \left\{\barr{l} 
 \wh{\gb}=\gb_0+\Ab^t(\Ab\Ab^t)^{+}(\db-\Ab\gb_0) \\ 
 \wh{\Rb}_g=\lambda(\Ib-\Ab^t(\Ab\Ab^t)^{+}\Ab)
 \earr\right..
\eeq
For the particular case of $\Ab=\cro{\stack{\Ab_1}{\Ab_2}}$ we have 
\[
\Ab\Ab^t
=\cro{\stack{\Ab_1\Ab_1^t}{\Ab_2\Ab_1^t}~\stack{\Ab_1\Ab_2^t}{\Ab_2\Ab_2^t}}
=\cro{\stack{K\Ib}{\oneb}~\stack{\oneb}{L\Ib}}, 
\]
where $\oneb$ is a matrix with all its elements equal to 1. 
We may note that  $\Ab\Ab^t$ is singular and its rank is $K+L-1$. 
We can however compute numerically $\wh{\gb}$ and $\wh{\Rb}_g$. 
Note also that, even if \aprio $g_{kl}$ were independent, \apost 
they are correlated. 

\bigskip\noindent{\bf Poisson case: $g_{kl}\sim \Pc(\lambda)$:}\\[6pt] 
Here, we have:
\beqnx
 P(g_{kl})=\lambda^{g_{kl}} \expf{-\lambda}/(g_{kl}!)\lra  
 \ln P(g_{kl})=(\ln\lambda)g_{kl}-\ln(g_{kl}!)-\lambda
\eeqnx
and 
\beqnx
 g_{kl}\sim\Pc(\lambda), \quad 
 r_k=\sum_l g_{kl}\sim\Pc(L\lambda), \quad 
 c_l=\sum_k g_{kl}\sim\Pc(K\lambda). 
\eeqnx
Then, we can write 
\beqnx
 P(\rb)=\prod_k (L\lambda)^{r_k} \expf{-L\lambda}/(r_{k}!), 
 \quad 
 P(\cb)=\prod_l (K\lambda)^{c_l} \expf{-K\lambda}/(c_{l}!)  
\eeqnx
and 
\beqnx
 P(g_{kl}|\rb,\cb,\lambda)&\propto&
 (\lambda)^{g_{kl}}/(g_{kl}!) 
 \prod_k (L\lambda)^{r_k}/(r_k!) 
 \prod_l (K\lambda)^{c_l}/(c_l!)\\ 
 & & \mbox{with~~~} r_k=\sum_l g_{kl}\mbox{~~and~~} c_l=\sum_k g_{kl}.
\eeqnx
It is then possible to show that $P(g_{kl}|\rb,\cb,\lambda)$ is also a 
Poisson law, but it is not easy to find an explicit expression for 
its mean value. However, using again the Striling formula when working 
with 
$\ln P(g_{kl}|\rb,\cb,\lambda)$ one can obtain an approximate expression 
for it  
\beqnx
 P(g_{kl}|\{g_{k'\not=k,l'\not=l}\},\rb,\cb,\lambda)=
 \Pc(\lambda(1+L\expf{c_l}+K\expf{r_k})), 
\eeqnx
and thus we have
\beqn
 \esp{g_{kl}|\{g_{k'\not=k,l'\not=l}\},\rb,\cb,\lambda}
 &=& \lambda(1+L\expf{c_l}+K\expf{r_k}) \nonumber \\ 
 &=& KL \lambda(1/(KL)+(1/K)\expf{c_l}+(1/L)\expf{r_k}).
 \nonumber \\  
\eeqn
This is interesting, because $(1/K)\expf{c_l}+(1/L)\expf{r_k}$ 
corresponds again to the famous back-projection operation in computed 
tomography, but here, in place of back-projecting $c_l$ and $r_k$ 
themselves, their exponential values $\expf{c_l}$ and $\expf{r_k}$ 
are back-projected. 

\section{Conclusions}

This paper was another analysis of dice problems trying to answer 
some of the questions about the situations where we can use the 
Bayesian or the Maximum Entropy approaches. 
Through this paper, we distinguished three approaches: 
Bayesian, classical MaxEnt and MaxEnt on the mean. 
I showed some of the situations where we can use these approaches. 

The Bayesian approach can be used 
when we can write explicitly a probabilistic model relating the 
data to the unknown parameters from which we can deduce 
the expression of the likelihood and can assign an 
\aprio law to those parameters, 
we can then use the Bayesian approach to compute the \apost 
from which we can infer about the parameters. 

The classical MaxEnt approach can be used in cases where we have a set 
of data which can be considered as linear constraints on a set of parameters 
which are themselves a probability distribution. 
Then the classical MaxEnt gives the possibility of  
finding a unique solution to the under-determined problem. 

The MaxEnt on the mean approach can be used in cases where we have a 
set of data which can be considered as linear constraints on the expected 
values of a set of parameters which are the elements of a convex set 
on which we can define a reference measure. 
Then, we can use the MaxEnt on the mean approach to compute a probability 
law on that set such that the expected values of the parameters satisfy 
exactly the data. 
We can then compute those expected values which depend on the choice of 
the reference measure. 
We showed also that there are strong relation between the two MaxEnt 
approaches.  

In some cases, it may happens that we have both the moment data and 
the sampling data. 
Then we can first use the MaxEnt approach 
to assign the prior law using the moment data and then use it with the 
likelihood to compute the \apost law of the parameters from which we can 
infer about them. 

Finally, even if I tried to answer to some of the questions, 
I also asked more questions to be answered. We thus still have a lot to 
do with all the three approaches. 
However, it seems that for practical applications the Bayesian approach 
seems to be the right and the easiest one. 

\bibliographystyle{ieeetr}
\def\bibpath{/home/djafari/Tex/Inputs/bib/}
\bibliography{bibenabr,revuedef,revueabr,baseAJ,baseKZ,\bibpath gpipubli,\bibpath amd1}

\begin{thebibliography}{10}

\bibitem{Jaynes68}
E.~T. Jaynes, ``Prior probabilities,'' {\em \uppercase{ieee} {T}rans. {S}ystems
  {S}cience and {C}ybernetics}, vol.~SSC-4, pp.~227--241, Sept. 1968.

\bibitem{Jaynes78}
E.~T. Jaynes, ``Where do we stand on maximum entropy ?,'' in {\em The Maximum
  Entropy Formalism} (R.~D. Levine and M.~Tribus, eds.), Cambridge (MA): M.I.T.
  Press, 1978.

\bibitem{Jaynes82}
E.~T. Jaynes, ``On the rationale of maximum-entropy methods,'' {\em {P}roc.
  \uppercase{ieee}}, vol.~70, pp.~939--952, Sept. 1982.

\bibitem{Jaynes85}
E.~T. Jaynes, ``Where do we go from here?,'' in {\em Maximum-Entropy and
  {B}ayesian Methods in Inverse Problems} (C.~R. Smith and W.~T.~J. Grandy,
  eds.), pp.~21--58, 1985.

\bibitem{Frieden85b}
Frieden, ``Dice, entropy and likelihood,'' {\em {J}. {O}pt. {S}oc. {A}mer.},
  vol.~73, no.~12, pp.~1764--1770, 1985.

\bibitem{Frieden87}
B.~R. Frieden, ``Maximum-probable restoration of photon-limited images,'' {\em
  {A}pplied {O}ptics}, vol.~26, no.~9, pp.~1755--1764, 1987.

\bibitem{Shore80}
J.~Shore and R.~Johnson, ``Axiomatic derivation of the principle of maximum
  entropy and the principle of minimum cross-entropy,'' {\em \uppercase{ieee}
  {T}rans. {I}nf. {T}heory}, vol.~26, pp.~26--37, Jan. 1980.

\bibitem{VanCampenhout81}
J.~M. Van~Campenhout and T.~M. Cover, ``Maximum entropy and conditional
  probability,'' {\em \uppercase{ieee} {T}rans. {I}nf. {T}heory}, vol.~27,
  pp.~483--489, July 1981.

\bibitem{Robert92a}
C.~Robert, {\em L'analyse statistique Bay\'esienne}.
\newblock Economica, 1992.

\bibitem{Robert97b}
C.~P. Robert, {\em The {B}ayesian Choice. {A} Decision-Theoretic Motivation}.
\newblock Springer Texts in Statistics, New York: Springer-Verlag, 1997.

\bibitem{Kullback51}
S.~Kullback and R.~A. Leibler, ``On information and sufficiency,'' {\em The
  Annals of Mathematical Statistics}, vol.~22, pp.~79--86, 1951.

\bibitem{Kullback59}
S.~Kullback, {\em Information Theory and Statistics}.
\newblock New York: Wiley, 1959.

\bibitem{Rockafellar70}
R.~T. Rockafellar, {\em Convex Analysis}.
\newblock Princeton University Press, 1970.

\bibitem{Borwein91a}
J.~M. Borwein and A.~S. Lewis, ``Duality relationships for entropy-like
  minimization problems,'' {\em \uppercase{siam} {J}. {C}ontrol and
  {O}ptimization}, vol.~29, pp.~325--338, Mar. 1991.

\bibitem{Djafari91a}
A.~Mohammad-Djafari, {\em A Matlab Program to Calculate the Maximum Entropy
  Distributions}, pp.~221--233.
\newblock Laramie, \sca{wy}: Kluwer Academic Publ., {T.W.}~{G}randy~ed., 1991.

\bibitem{Rockafellar93}
R.~T. Rockafellar, ``Lagrange multipliers and optimality,'' {\em SIAM Review},
  vol.~35, pp.~183--238, June 1993.

\bibitem{LeBesnerais93a}
G.~Le~Besnerais, {\em M\'ethode du maximum d'entropie sur la moyenne,
  crit\`eres de reconstruction d'image et synth\`ese d'ouverture en
  radio-astronomie}.
\newblock Phd thesis, Universit\'e de Paris-Sud, Orsay, Dec. 1993.

\bibitem{Djafari94}
A.~Mohammad-Djafari, ``Maximum d'entropie et probl\`emes inverses en
  imagerie,'' {\em {T}raitement du {S}ignal}, pp.~87--116, 1994.

\bibitem{Bercher95a}
J.-F. Bercher, {\em D\'eveloppement de crit\`eres de nature entropique pour la
  r\'esolution des probl\`emes inverses lin\'eaires}.
\newblock Phd thesis, Universit\'e de Paris--Sud, Orsay, Feb. 1995.

\bibitem{Djafari96g}
A.~Mohammad-Djafari, ``A comparison of two approaches: {M}aximum entropy on the
  mean ({MEM}) and {B}ayesian estimation ({BAYES}) for inverse problems,'' in
  {\em {M}aximum {E}ntropy and {B}ayesian {M}ethods}, (Berg--en--Dal,),
  {K}luwer {A}cademic {P}ubl., Aug. 1996.

\bibitem{LeBesnerais99}
G.~Le~Besnerais, J.-F. Bercher, and G.~Demoment, ``A new look at entropy for
  solving linear inverse problems,'' {\em \uppercase{ieee} {T}rans. {I}nf.
  {T}heory}, vol.~45, pp.~1565--1578, July 1999.

\bibitem{Djafari99b}
A.~Mohammad-Djafari, ``Entropie en traitement du signal,'' {\em Traitement du
  signal}, vol.~Num. sp\'ecial, volume 15, no.~6, pp.~545--551, 1999.

\end{thebibliography}

\end{document}